\documentclass[conference]{IEEEtran}
\usepackage[utf8]{inputenc}
\usepackage[utf8]{inputenc}
\usepackage[english]{babel}
\usepackage[T1]{fontenc}
\usepackage{amsmath}
\usepackage{amsfonts}
\usepackage{amssymb}
\usepackage{url}
\usepackage[official]{eurosym}
\usepackage{pgfplots}
\pgfplotsset{width=7cm}
\usetikzlibrary{patterns}
\usepackage{tikz-cd}
\usetikzlibrary{shapes,arrows}
\usetikzlibrary{calc,fit,trees,positioning,arrows,chains,shapes.geometric,shapes}
\usetikzlibrary{shapes,arrows}
\author{\IEEEauthorblockN{Eduardo Rodrigues\IEEEauthorrefmark{1}, Ricardo Morla\IEEEauthorrefmark{2}}
\IEEEauthorblockA{ INESC TEC\IEEEauthorrefmark{2}, Faculty of Engineering, University of Porto\\Porto, Portugal \\
Email: \IEEEauthorrefmark{1}up201304756@fe.up.pt, \IEEEauthorrefmark{2}ricardo.morla@fe.up.pt
}}

\usepackage{listings}

\title{Run Time Prediction for Big Data Iterative ML Algorithms: a  KMeans case study}

\begin{document}

\maketitle

\begin{abstract}
Data science and machine learning algorithms running on big data infrastructure are increasingly important in activities ranging from business intelligence and analytics to cybersecurity, smart city management, and many fields of science and engineering. 
As these algorithms are further integrated into daily operations, understanding how long they take to run on a big data infrastructure is paramount to controlling costs and delivery times. In this paper we discuss the issues involved in understanding the run time of iterative machine learning algorithms and provide a case study of such an algorithm - including a statistical characterization and model of the run time of an implementation of K-Means for the Spark big data engine using the Edward probabilistic programming language.

\end{abstract}

\section{Introduction}
\label{sec:intro}

The time an iterative machine learning algorithm takes to run on a big data infrastructure is not likely to be the same at every run. The factors involved can be non-deterministic and related to 1) data -- namely the size and distribution of the data, 2) the algorithm -- how complex each iteration is, how fast the algorithm converges, how the algorithm is implemented, any pseudo-random seeds the algorithm is likely to use, 3) the infrastructure resources available -- how many nodes the big data cluster has, how many cores and RAM are available per node, node CPU and I/O speed, 4) the big data software management stack that is used and how it randomly schedules infrastructure resources to the application, and finally 5) load on the infrastructure that is external to the application -- how many resources are shared with other applications and IT services in the cluster and how that affects the performance of the iterative ML application whose run time we are interested in.

Predicting the run time of an iterative machine learning algorithm may be interesting in different circumstances. For example you may have a large data set you need to process and want to have a better idea of how long your infrastructure will take to process the data by running the algorithm on a sample of the data first. You may also want to understand how much you need to grow your infrastructure to get results faster, how long it will take to process more data, or how much longer will the currently running algorithm take to produce results. 

Researchers have been looking at this problem through different perspectives throughout the years. A single data-intensive job greedily scheduled on a big data cluster has minimum and maximum  duration specified by the Makespan theorem, and this has been used specifically to improve resource provisioning of MapReduce jobs~\cite{verma_resource_2011} and more recently for general-purpose Spark applications~\cite{wang_performance_2015}. Predicting run time specifically for large-scale iterative machine learning applications has mostly focused on sampling from the original data set: \cite{venkataraman_ernest_2016} uses non-negative least-squares to select best set of features to predict the run time of the full data set and evaluates performance on a reasonably large number of algorithms, whereas \cite{popescu_predict_2013} focuses on graph data and algorithms and specifically attempts to predict the number of algorithm iterations. Finally, researchers have also been looking at theoretical bounds for iterative algorithms. Specifically for KMeans a proof is given for an exponential lower bound on the number of data points for its run time  \cite{vattani_k-means_2009}.

In this paper we set out to design a probabilistic model for large scale iterative machine learning algorithms, which previous work has not covered. The first step of this design is bootstrapped by the KMeans algorithm. In the rest of the paper we describe the model and its limitations and provide some results that show variability of the run time and how it can be used to improve run time estimate during application run.

\section{Approach}

Spark\footnote{\url{https://spark.apache.org}} defines a clear hierarchy for its applications. Applications are divided into jobs. Jobs run serially if they are triggered by the same thread in the Spark master application or in parallel if they are triggered by different threads in the master. Each job has a sequence of stages, each of which spawns a set of tasks that run concurrently in the different cores and nodes of the Spark cluster. A stage is only over when all its tasks are over. 

Our approach uses the first two levels of this hierarchy -- application and job -- to model application run time from job run time. The KMeans implementation\footnote{\url{https://github.com/apache/spark/blob/master/examples/src/main/python/mllib/kmeans.py}} we use is based on kmeans||\footnote{\url{http://theory.stanford.edu/~sergei/papers/vldb12-kmpar.pdf}} and its jobs can be divided into two types: a set of non-iterative jobs that performs initialization of the cluster centers and algorithm finalization and a set of iterative jobs that possibly converges to the final cluster centers. We use normal variables to model iterative and non-iterative jobs. Mean and variance for iterative job variables are computed from the run time of all iterative jobs, while for non-iterative jobs only the run time data corresponding to the exact job is used. 

We use the python-based Edward\footnote{\url{https://edwardlib.org}} Turing-complete probabilistic programming language to define the model for application run time and perform inference. The application run time is the sum of all job run time plus any remaining time that Spark may take to spawn jobs. In addition to the job run time random variables, we define a random variable for the number of iterations of the KMeans algorithm. We  express our model as follows: $A = \sum J_{ni} + \sum_{i=1}^{i=NI} J_i$ where $NI$ is the random variable for the number of iterations, $J_i$ and $J_{ni}$ are the random variables for the different iterative and non-iterative jobs, and $A$ is the random variable for the application run time. In this paper we assume $NI$ is known and fixed and expect to explore modeling the number of iterations in future work.

We start by defining the iterative jobs as independent variables. To account for factors that may affect one particular run of the application -- such as temporary additional load on the infrastructure while one instance of the application is running -- we also define iterative jobs $J_i$ as dependent normal variables. In this case the mean of $J_i$ is defined by another normal variable $A_{mean}$ as follows: $J_i \sim \mathcal{N}(\mu=A_{mean},\sigma^2)$. We compute mean and variance of $A_{mean}$ using the per application average run time of the application's iterative jobs. Non-iterative jobs are modeled as independent variables as before.

Listing \ref{lst:code_direct} shows part of the source code that we used to implement our approach. A list of random variables is passed to the application run time function $A$, which is then used for inference. Notice the mean of the independent normal variables is defined as a real number whereas the mean of the dependent normal variable is defined as variable $A_{mean}$. Application and job run time data used for estimating parameters are extracted directly from the log files available from Spark's master node.

\lstset{basicstyle=\footnotesize}
\lstset{language=Python}
\lstset{frame=lines}
\lstset{belowcaptionskip=1cm}
\lstset{caption={Code Snippet for Independent and Dependent Approaches}}
\lstset{label={lst:code_direct}}
\begin{lstlisting}[frame=single]

from edward.models import Normal
# app_mu, app_sc, job_mu, job_sc: 
# floats, estimated from data

# Independent:
job_ind = []
for k in iterjobs:
	job_ind.append(\
	Normal(loc=job_mu, scale=job_sc))

# Dependent:
A_mean = Normal(loc=app_mu, scale=app_sc)
job_dep = []
for k in iterjobs:
	job_dep.append(\
	Normal(loc=A_mean, scale=job_sc))


# Application run time function
def A(joblist):
    t = 0
    for J_i in joblist:
        t = t + J_i
    return t
\end{lstlisting}

\section{Results}

We deployed the algorithm on a Spark cluster running on Chameleon cloud\footnote{\url{https://chameleoncloud.org}} with 1 master node and 3 worker nodes. Each worker node has 48 cores and 128GB of RAM. We use this data\footnote{\url{http://archive.ics.uci.edu/ml/datasets/Individual+household+electric+power+consumption}} from the UCI Machine Learning Repository for clustering. We ignore the date and time fields and replicate the data to achieve the desired file size of 10GB. The number of clusters of the KMeans algorithm was set to 4. In order to avoid as much as possible randomness from the algorithm and the data we decided to use the same data and random seed in all the runs of KMeans in this paper. Given the exploratory nature of this paper we do not show cross-validated results.

\begin{figure}[h!]
\begin{center}
\includegraphics[width=0.5\textwidth]{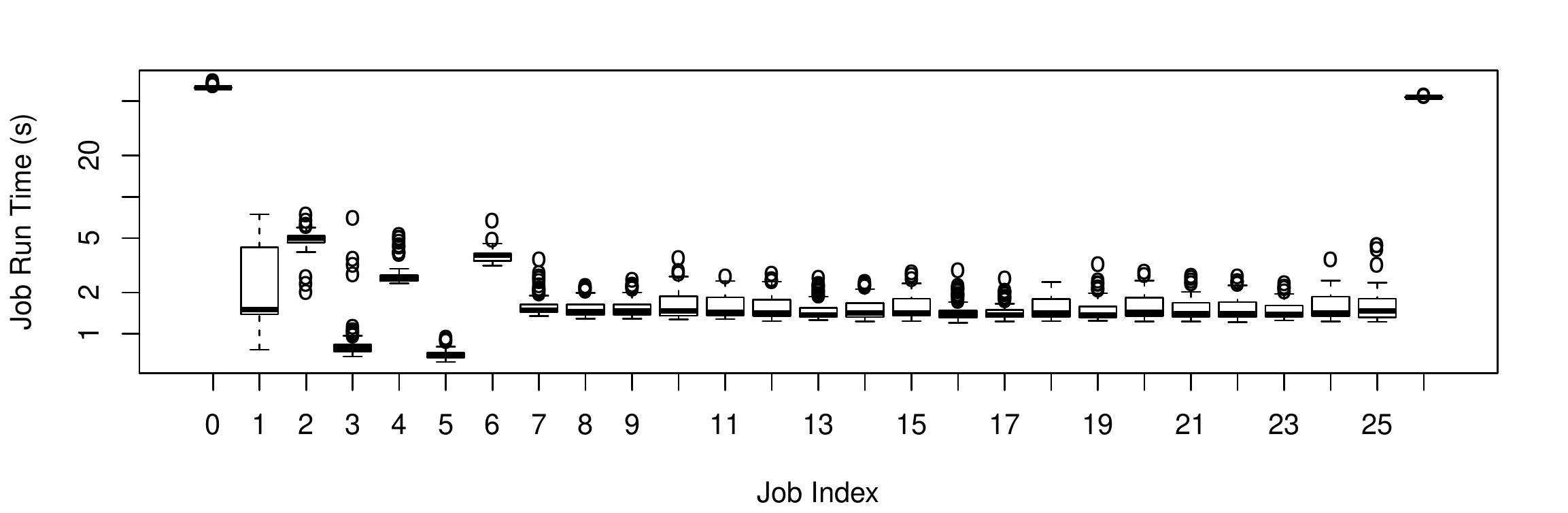} 
\end{center}
\caption{Boxplot diagrams for job run time per job index. Jobs 7-25 are iterative.}
\label{runtimeperjobindex}
\end{figure}

\begin{figure}[h!]
\begin{center}
\includegraphics[width=0.24 \textwidth]{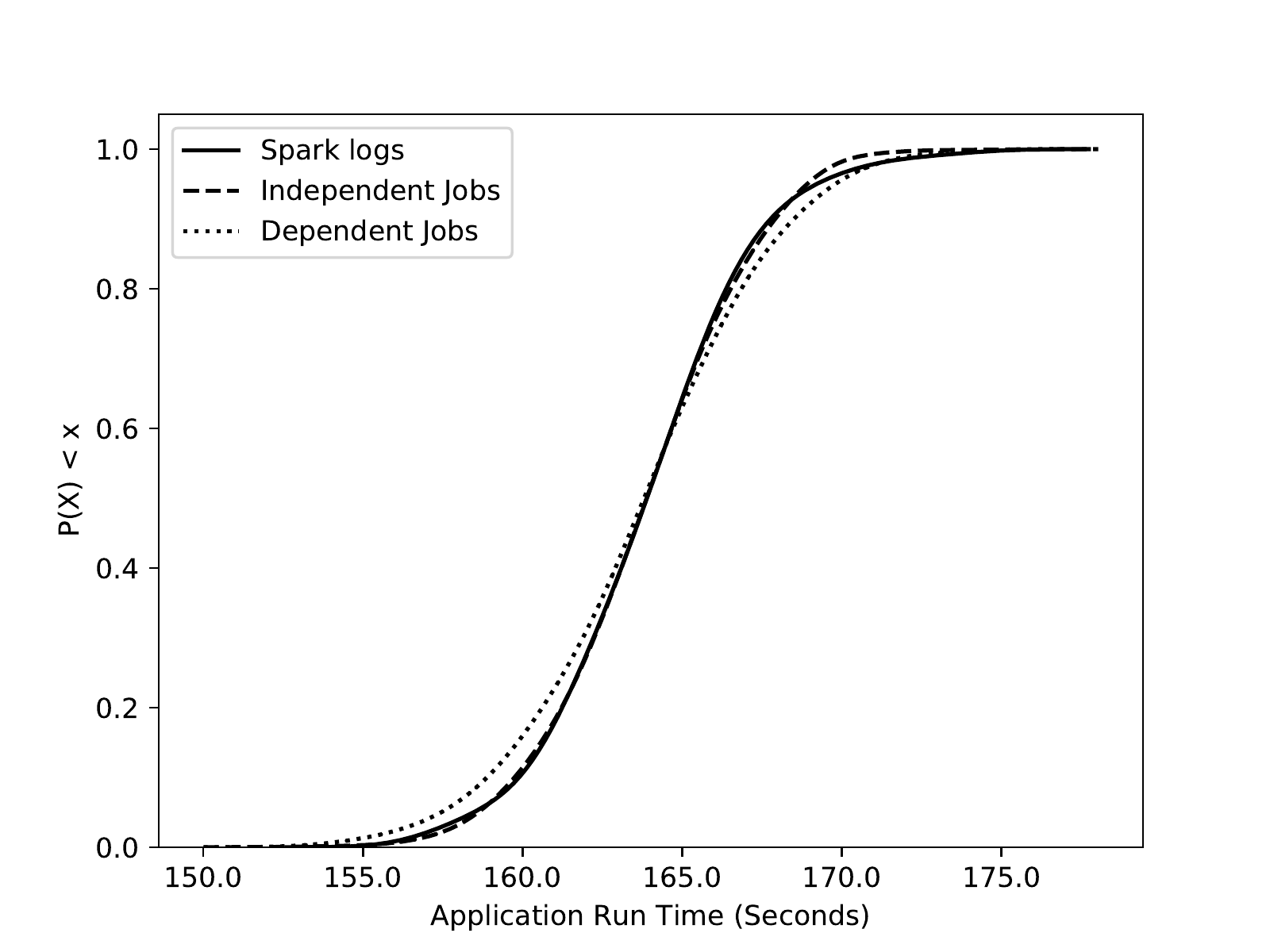} 
\includegraphics[width=0.24 \textwidth]{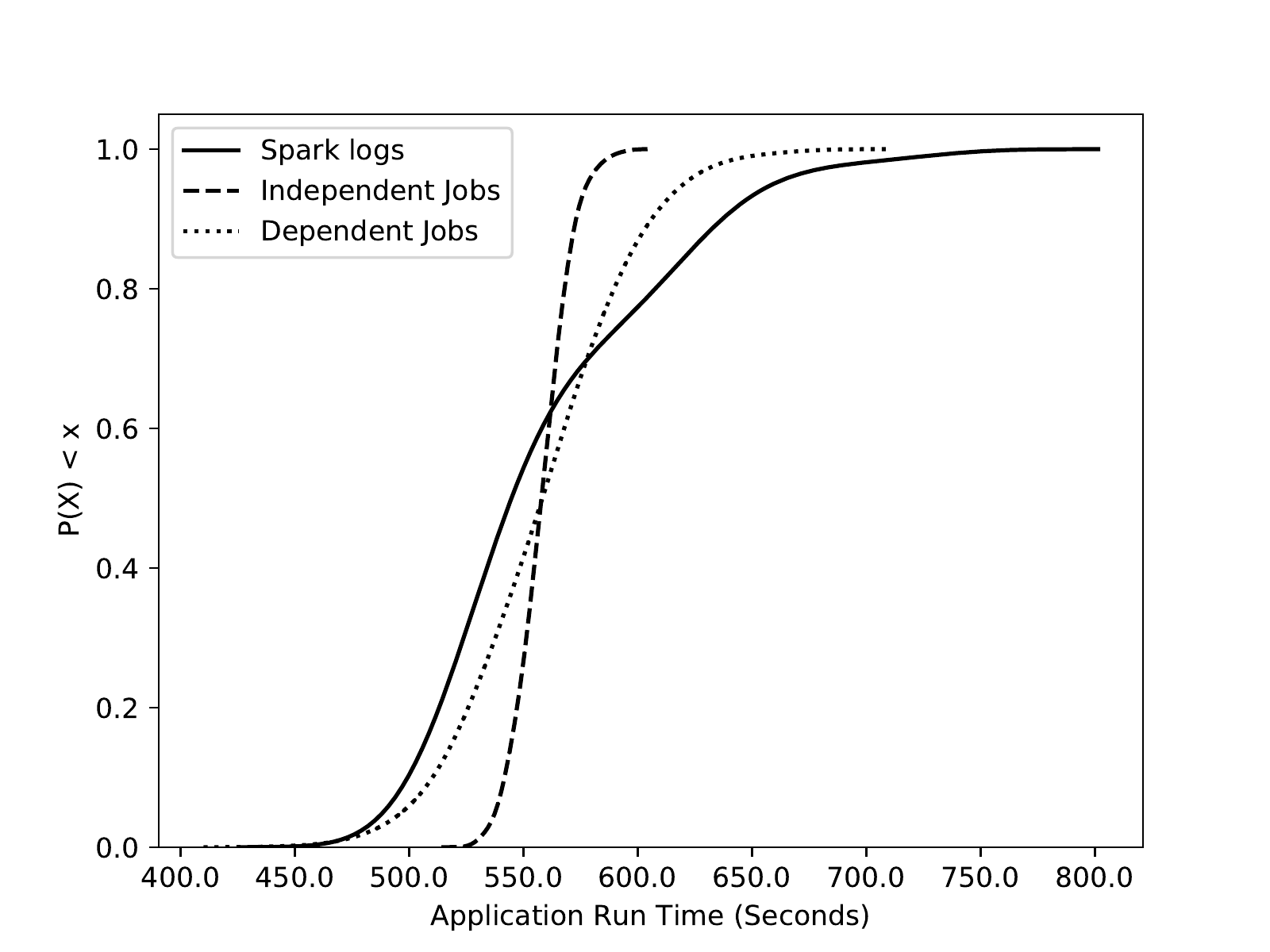} 
\end{center}
\caption{Empirical Cumulative Distribution Function of Application Run Time - actual (Spark logs) and predicted using independent and dependent models for job run time. Left: Spark workers use total memory available in the node. Right: Each Spark worker only use 1GB RAM.}
\label{paper-all}
\end{figure}

We ran the KMeans algorithm 100 times and allowed it to access the full memory of each node on the Spark cluster. This experiment resulted in an application run time median of 164.0 s with $1^{st}$ and $99^{th}$ percentiles at 157.1 s and 171.4 s respectively. Figure \ref{runtimeperjobindex} shows the boxplot of job run time for each job index in the application. As expected given that we are using the same data and random seed, the number of iterations in the algorithm -- and as such the number of jobs per application run -- is the same. Figure \ref{paper-all} (left) shows the results of applying our models in this experiment. Both dependent and independent approaches behave particularly well with a visibly strong overlap of resulting distributions.

We conducted another experiment that restricts the amount of RAM that can be accessed by the application in each node to 1GB. Limiting the memory to 1GB results in the garbage collector having to run for much longer than in the other experiment. This difference is especially large since in the other experiment there are 128 GB RAM available to the application which is more than sufficient for the garbage collector to almost never be invoked. Moreover, given the seemingly random nature of the garbage collection mechanism, some applications are expected to experience delay because of garbage collection and others are not. Designing a system to use the maximum memory available is not economically viable and garbage collection is used extensively in actual deployments. Figure \ref{paper-all} (right) shows the results of applying our models in this experiment. Neither dependent nor independent models match the actual application run time distribution as precisely as in the other experiment. However it is possible to observe that the dependent model fits the application run time distribution better than the independent model and that the independent model yields a significantly narrower range of application run time values than that of the actual run time distribution. One way of improving the results for the dependent model could to use a mixture model for $A_{mean}$ instead of a normal variable.

\begin{figure}[h!]
\begin{center}
\includegraphics[width=0.24 \textwidth]{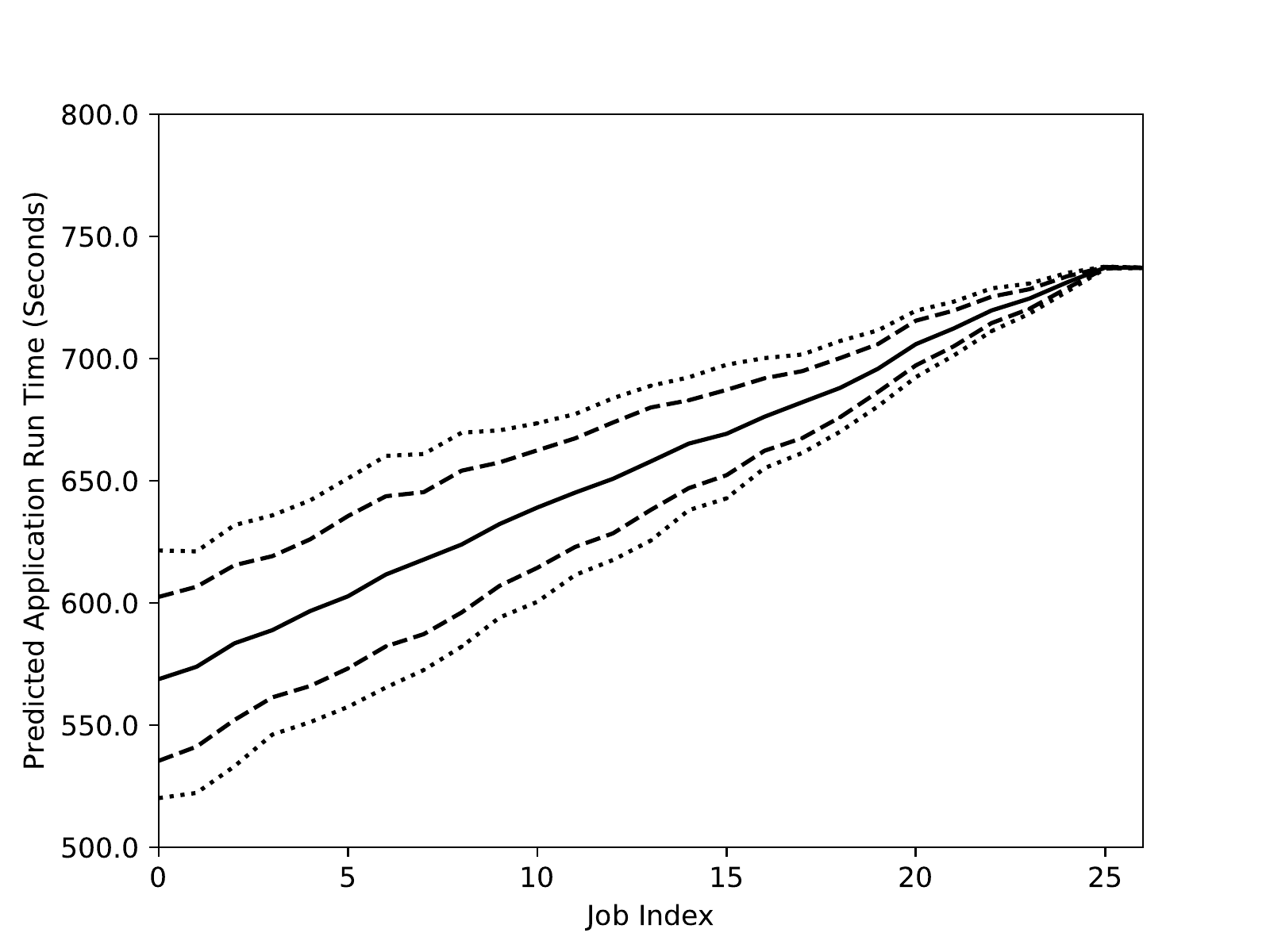} 
\includegraphics[width=0.24 \textwidth]{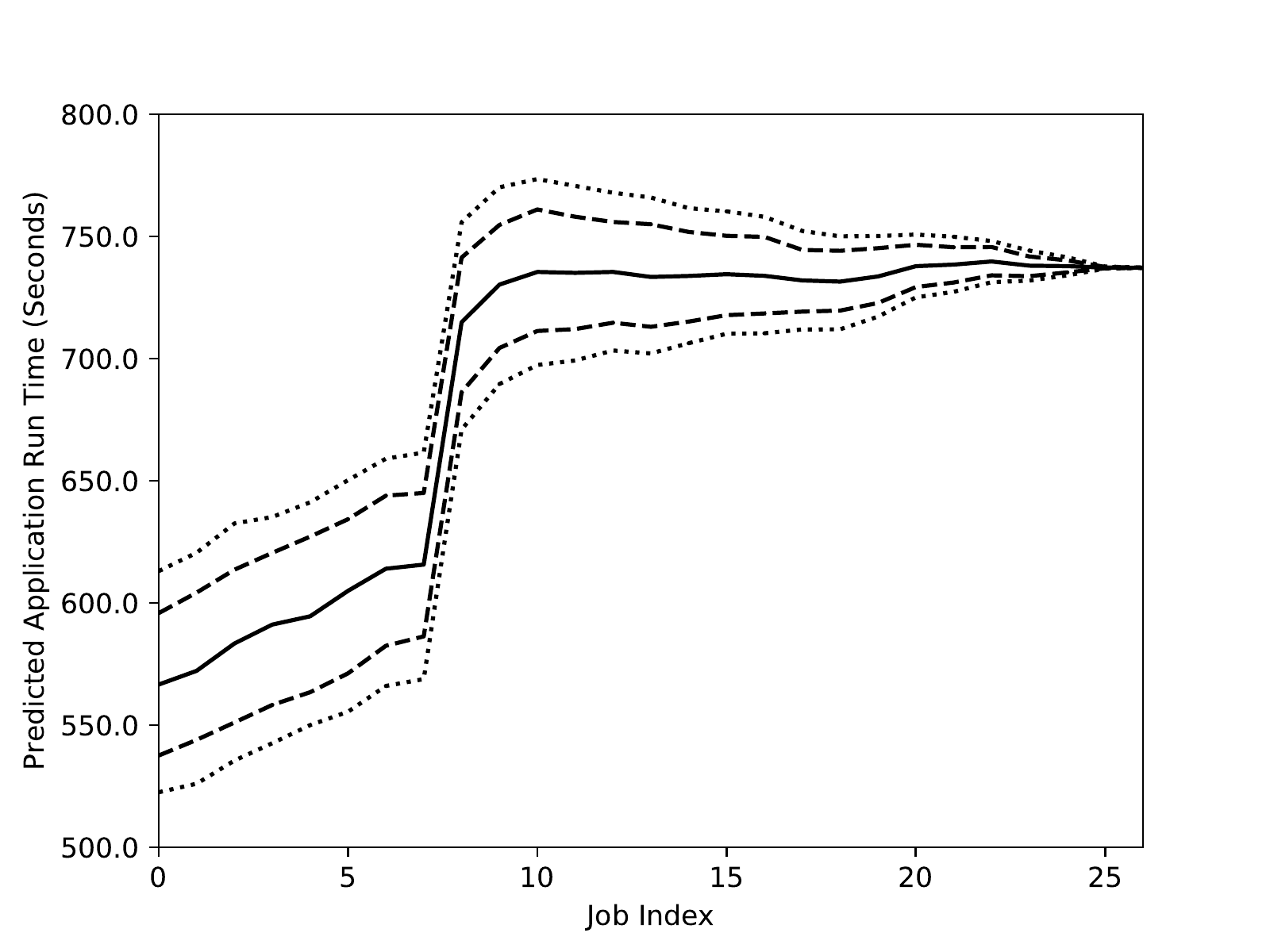} 
\end{center}
\caption{Evolution of the run time prediction of an application using the dependent variable approach on the 1GB RAM experiment and the longest application run on this experiment. Solid: percentile 50, dashed: percentiles 20 and 80, dotted: percentiles 10 and 90. Left: using the same $A_{mean}$ as before. Right: updating $A_{mean}$ after the first iterative job with the average of iterative jobs observed so far on this run.  }
\label{paper-evolv}
\end{figure}

Finally we provide an example of how this modeling approach can be used to improve estimates of the currently running application run time as its jobs gradually finish. At the end of each job, we would like to update our initial distribution of application run time. Simply subtracting the elapsed run time from the initial distribution does not satisfy us -- it will eventually produce negative results and  run time variance will not be reduced, which should happen as we remove the uncertainty in the run time of the jobs that have finished. Our approach involves inference of the application run time at the end of every job. We do so by removing finished jobs from the list of job run time random variables that are passed as arguments to the application run time function $A$ and resampling A at the end of each job. This results in Figure \ref{paper-evolv} (left) where we can see that the percentile lines are approaching and narrowing in on the actual application run time as the job index increases.

The results shown in Figure \ref{paper-evolv} (left) are from the longest application run of the experiment. The actual application run time is outside the 20 and 10 percentile lines almost until the very last jobs. In order to improve this, we consider another approach where the mean of $A_{mean}$ is updated to be the average of the observed run times of all finished iterative jobs if any of these jobs has finished. Figure \ref{paper-evolv} (right) shows the results of this approach. After the first iterative job finishes (job 7) the prediction quickly readjusts to the newly observed iterative job run time and afterwards remains centered and narrows in on the actual application run time.

\section{Conclusion}

We hope to extend the KMeans work by modeling the number of iterations and varying data size. We also intend to understand how our model works with other iterative ML algorithms. 

\section*{Acknowledgments}
Results presented in this paper were obtained using the Chameleon testbed supported by the National Science Foundation.

\bibliographystyle{unsrt}
\bibliography{arxiv}
\end{document}